\documentclass[twocolumn,superscriptaddress,showpacs,floatfix,aps,prb]{revtex4}
\usepackage{graphicx}
\usepackage{units}
\usepackage{enumerate} 
\usepackage{bm}
\usepackage{amsmath}
\usepackage{amsfonts}
\usepackage{amssymb}
\usepackage[thinspace,squaren,pstricks,derivedinbase,derived]{SIunits}
\usepackage{CJKutf8}
\usepackage[pdfpagelabels, colorlinks, citecolor=blue]{hyperref}

\begin{document}

\title{Propagating spin waves in nanometer-thick yttrium iron garnet films: Dependence on wave vector, magnetic field strength and angle}

\author{Huajun Qin} 
\email{huajun.qin@aalto.fi}
\affiliation {NanoSpin, Department of Applied Physics, Aalto University School of Science, P.O. Box 15100, FI-00076 Aalto, Finland}
\author{Sampo J. Hämäläinen} \affiliation {NanoSpin, Department of Applied Physics, Aalto University School of Science, P.O. Box 15100, FI-00076 Aalto, Finland}
\author{Kristian Arjas} \affiliation {NanoSpin, Department of Applied Physics, Aalto University School of Science, P.O. Box 15100, FI-00076 Aalto, Finland}
\author{Jorn Witteveen} \affiliation {NanoSpin, Department of Applied Physics, Aalto University School of Science, P.O. Box 15100, FI-00076 Aalto, Finland}
\author{Sebastiaan van Dijken}
\email{sebastiaan.van.dijken@aalto.fi}
\affiliation {NanoSpin, Department of Applied Physics, Aalto University School of Science, P.O. Box 15100, FI-00076 Aalto, Finland}

\begin{abstract}
We present a comprehensive investigation of propagating spin waves in nanometer-thick yttrium iron garnet (YIG) films. We use broadband spin-wave spectroscopy with integrated coplanar waveguides (CPWs) and microstrip antennas on top of continuous and patterned YIG films to characterize spin waves with wave vectors up to 10 rad/$\mu$m. All films are grown by pulsed laser deposition. From spin-wave transmission spectra, parameters such as the Gilbert damping constant, spin-wave dispersion relation, group velocity, relaxation time, and decay length are derived and their dependence on magnetic bias field strength and angle is systematically gauged. For a 40-nm-thick YIG film, we obtain a damping constant of $3.5 \times 10^{-4}$ and a maximum decay length of 1.2 mm. Our experiments reveal a strong variation of spin-wave parameters with magnetic bias field and wave vector. Spin-wave properties change considerably up to a magnetic bias field of about 30 mT and above a field angle of $\theta_{H} = 20^{\circ}$, where $\theta_{H} = 0^{\circ}$ corresponds to the Damon-Eshbach configuration.  
\end{abstract}

\date{\today}
\maketitle

\section{Introduction}
Magnonics aims at the exploitation of spin waves for information transport, storage, and processing \cite {Serga2010, Kruglyak2010, Lenk2011, Krawczyk2014, Chumak2015, Nikitov2015, Chumak2017}. For practical devices, it is essential that spin waves propagate over long distances in thin films. Because of its ultralow damping constant, ferrimagnetic YIG is a promising material. Bulk crystals and $\mu$m-thick YIG films exhibit a Gilbert damping constant $\alpha \approx 3 \times 10^{-5}$ at GHz frequencies. In recent years, nm-thick YIG films with ultralow damping parameters have also been prepared successfully. High-quality YIG films have been grown on Gd$_{3}$Ga$_{5}$O$_{12}$ (GGG) single-crystal substrates using liquid phase epitaxy \cite {Hahn2013, Pirro2014, Wang2016, Dubs2017}, magnetron sputtering \cite {Chang2014, Lustikova2014, Gallagher2016, CaoVan2018}, and pulsed laser deposition (PLD) \cite {Heinrich2011, Sun2012, Kelly2013, Onbasli2014, Howe2015, Lutsev2016, Hauser2016, Sokolov2016, Krysztofik2017, Bhoi2018}. For thin YIG films, damping constants approaching the value of bulk crystals have been reported \cite {Hauser2016, Lutsev2016}. Meanwhile, YIG-based magnonic devices such as logic gates, transistors, and multiplexers have been demonstrated \cite {Schneider2008, Ganzhorn2016, Fischer2017, Chumak2014, Davies2015}. Spin-wave transmission in nm-thick YIG films \cite{Yu2014, Khivintsev2015, Collet2017, Krysztofik2017, Maendl2017, Chen2018, Liu2018} and the excitation of short-wavelength spin waves have been investigated as well \cite{Yu2016, Qin2018, Klingler2018, Maendl2018}. To advance YIG-magnonics further, knowledge on the transport of spin waves in nm-thick YIG films and its dependence on wave vector and external magnetic bias field is essential. 

In this paper, we present a broadband spin-wave spectroscopy study of PLD-grown YIG films with a thickness of 35 nm and 40 nm. Spin-wave transmission spectra are recorded by patterning CPWs and microstrip antennas on top of continuous and patterned YIG films. CPWs are used because they generate spin waves with well-defined wave vectors. This enables extraction of key parameters such as the Gilbert damping constant ($\alpha$), the spin-wave dispersion relation, group velocity ($\upsilon_{g}$), relaxation time ($\tau$), and decay length ($l_{d}$). For a 40 nm YIG film, we find $\alpha \approx 3.5 \times 10^{-4}$ and a maximum group velocity and decay length of 3.0 km/s and 1.2 mm, respectively. We show that spin-wave properties vary strongly with wave vector up to an in-plane external magnetic bias field $\mu_{0} H_{ext} = 30$ mT and below a field angle $\theta_{H} = 20^{\circ}$ ($\theta_{H}$ = 0 corresponds to the Damon-Eshbach geometry). Beyond these field parameters, the dependence of spin-wave properties on wave vector weakens. We demonstrate also that broadband spectroscopy with integrated CPWs and microstrip antennas provide similar spin-wave parameters.     

The paper is organized as follows. In Sec. II, we describe the PLD process, broadband spin-wave spectroscopy setup, and simulations of the CPW- and microstrip-antenna excitation spectra. In Sec. III, we present vector network analyzer ferromagnetic resonance (VNA-FMR) results and broadband spin-wave transmission spectra for CPWs. In Sec. IV, we fit the experimental data and extract parameters of propagating spin waves. Spin-wave transmission measurements using CPWs and microstrip antennas are compared in Sec. V. Section VI summarizes the paper.                   

\section{Experiment}
\subsection{PLD of YIG thin films} 
YIG films with a thickness of 35 nm and 40 nm were grown on single-crystal GGG(111) substrates using PLD. Prior to loading into the PLD vacuum chamber, the substrates were ultrasonically cleaned in acetone, isopropanol, and distilled water. The substrates were first degassed at $550^{\circ}$C for 15 minutes and then heated to $800^{\circ}$C at a rate of $5^{\circ}$C per minute in an O$_{2}$ pressure of 0.13 mbar. YIG films were deposited under these conditions by ablation from a stoichiometric target using an excimer laser with a pulse repetition rate of 2 Hz and a fluence of 1.8 J/cm$^{2}$. After deposition, the YIG films were first annealed at $730^{\circ}$C for 10 minutes in 13 mbar O$_{2}$ before cooling down to room temperature at a rate of $-3^{\circ}$C per minute. 

\subsection{Structural and magnetic characterization}
\begin{figure}[t]
	\vspace{10pt} 
	\resizebox*{0.98\columnwidth}{!}{\includegraphics{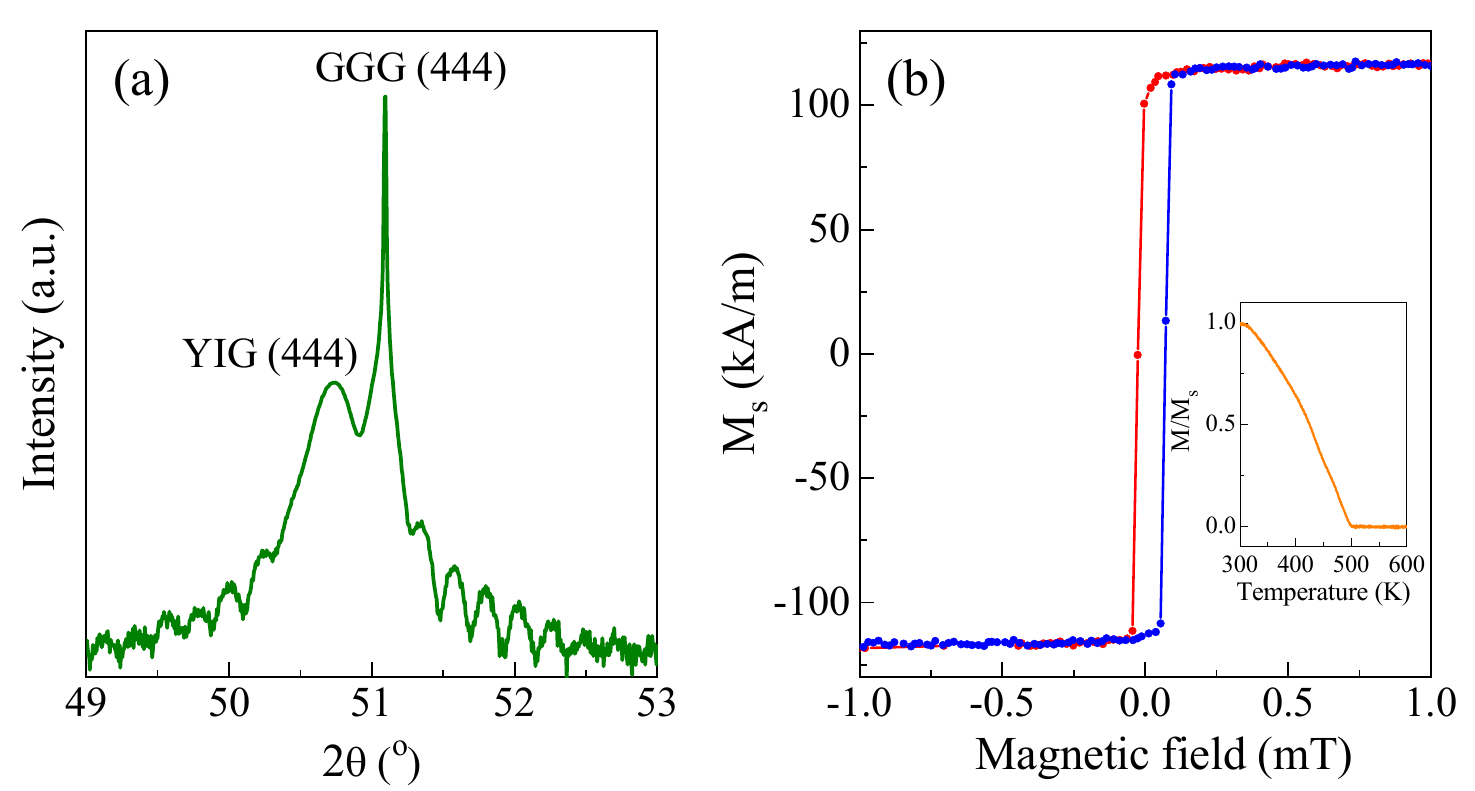}}
	\caption{(a) XRD $\theta-2\theta$ scan of the (444) reflections from a PLD-grown YIG film on a GGG(111) substrate. The period of Laue oscillations surrounding the (444) peaks corresponds to a film thickness of 40 nm. (b) Room temperature VSM hysteresis loop of the same film. The inset shows how the YIG saturation magnetization varies with temperature.}
	\label{Fig1}
\end{figure}

The crystal structure of our YIG films was inspected by high-resolution X-ray diffraction (XRD) on a Rigaku SmartLab system. Figure \ref{Fig1}(a) shows a XRD $\theta-2\theta$ scan of a 40-nm-thick YIG film on GGG(111). Clear (444) film and substrate peaks are surrounded by Laue oscillations, signifying epitaxial and smooth film growth. We used a vibrating sample magnetometer (VSM) in a PPMS Dynacool system from Quantum Design to characterize the magnetic properties. Figure \ref{Fig1}(b) depicts a VSM hysteresis loop of a 40-nm-thick YIG film. At room temperature, the coercive field of the YIG film is only 0.1 mT and the saturation magnetization ($M_{s}$) is 115 kA/m. The evolution of $M_{s}$ with temperature is shown in the inset of Fig. \ref{Fig1}(b). From these data, we derive a Curie temperature ($T_{C}$) of around 500 K. The values of $M_{s}$ and $T_{C}$ are similar to those obtained in previous studies on nm-thick YIG films \cite{Sokolov2016, Hauser2016, Gallagher2016} and about $10\%$ smaller compared to values of YIG bulk crystals ($M_{s}$ = 139 kA/m, $T_{C}$ = 559 K). Minor off-stoichiometries in the YIG film might be the reason for the small discrepancy \cite {Krockenberger2008}. 

\subsection{Broadband spin-wave spectroscopy} 
VNA-FMR and spin-wave transmission measurements were performed using a two-port VNA and a microwave probing station with a quadrupole electromagnet. In VNA-FMR experiments, the YIG film was placed face-down onto a prepatterned CPW on a GaAs substrate. The signal line and ground lines of this CPW had a width of 50 $\mu$m and 800 $\mu$m, respectively, and were separated by 30 $\mu$m. Broadband spin-wave spectroscopy in transmission geometry was conducted by contacting two integrated CPWs or microstrip antennas on top of a continuous YIG film or YIG waveguide. Most of the experiments were performed with CPWs consisting of 2 $\mu$m-wide signal and ground lines with a separation of 1.6 $\mu$m. For comparison measurements, we used CPWs and microstrip antennas with 4-$\mu$m-wide signal lines. All antenna structures were fabricated by electron-beam lithography and were composed of 3-nm Ta and 120-nm Au. A microwave current provided by the VNA was used to generate a rf magnetic field around one of the CPWs or microstrip antennas. We used CST microwave studio software to simulate the excitation spectra of the antenna structures (see next section). 

Spin waves that are excited by a rf magnetic field produce an inductive voltage across a nearby antenna. At the exciting CPW or microstrip antenna, this voltage is given by \cite {Huber2013}:
\begin{equation}\label{Eq:FMR}
V_{ind} \propto \int {\chi(\omega, k)} |\rho(k)|^{2}dk,
\end{equation}
where $\chi(\omega, k)$ is the magnetic susceptibility and $|\rho(k)|^{2}$ is the spin-wave excitation spectrum. Propagating spin waves arriving at the receiving CPW or microstrip antenna produce an inductive voltage:
\begin{equation}\label{Eq:PSW}
V_{ind} \propto \int {\chi(\omega, k)} |\rho(k)|^{2} \exp(-i(ks+\Phi_{0}))dk,
\end{equation}
where $s$ is the propagation distance and $\Phi_{0}$ is the initial phase of the spin waves. In the experiments, we used the first and second port of the VNA to measure these inductive voltages by recording the $S_{12}$ scattering parameter.

\begin{figure*}[t]
	\vspace{10pt} 
	\resizebox*{1.8\columnwidth}{!}{\includegraphics{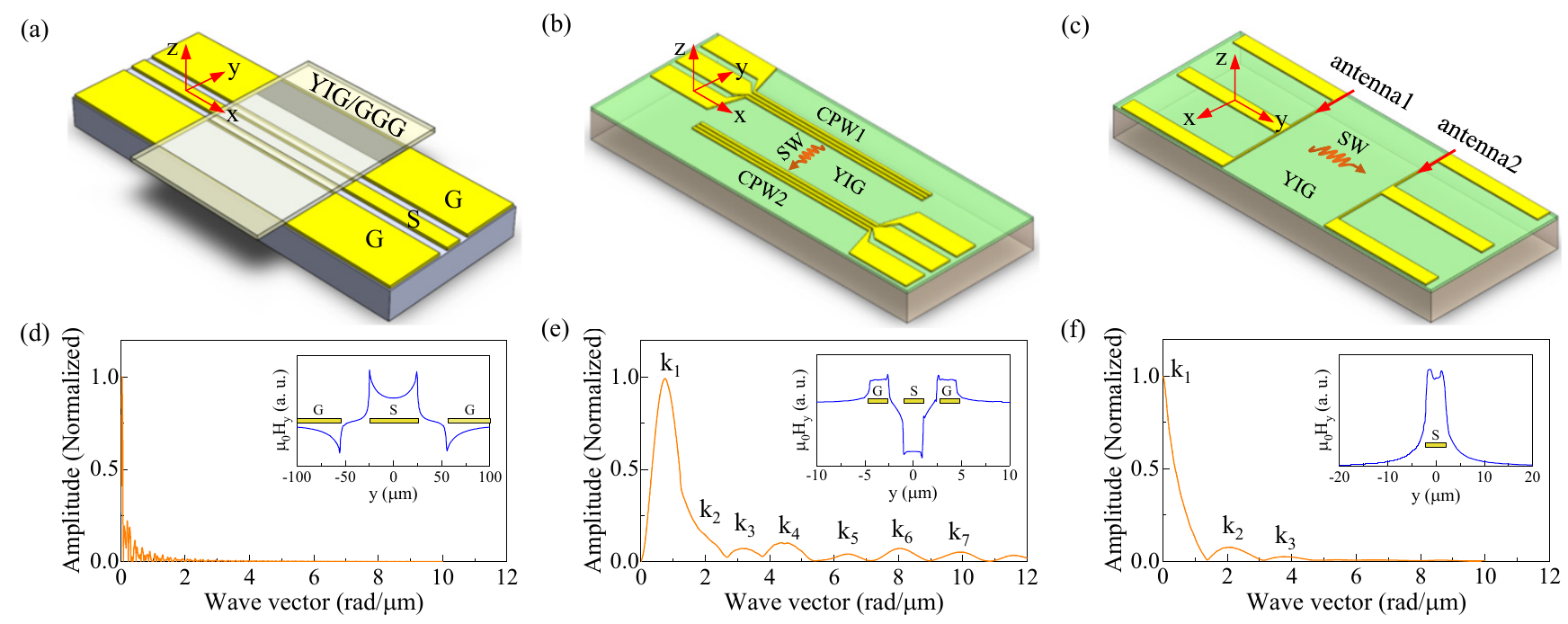}}
	\caption{(a-c) Schematic illustrations of several measurement configurations used in this study. (a) VNA-FMR measurements are performed by placing the YIG/GGG sample face-down onto a CPW. The CPW consists of a 50 $\mu$m-wide signal line and two 800 $\mu$m-wide ground lines. The gap between the signal and ground lines is 30 $\mu$m. (b-c) Spin-wave transmission through the YIG film is characterized by patterning two CPWs (b) or two microstrip antennas (c) on top of a YIG film. The signal and ground lines of the CPWs in (b) are 2 $\mu$m wide and separated by 1.6 $\mu$m gaps. The microstrip antennas, which are marked by red arrows in (c), are 4 $\mu$m wide. (d-f) Simulated spin-wave excitation spectra of the different antenna structures. The in-plane rf magnetic fields ($\mu_{0} h^{rf}_{y}$) that are produced by passing a microwave current through the CPWs in (a) and (b) or the microstrip antenna in (c) are shown in the insets.}
	\label{Fig2}
\end{figure*} 

\subsection{Simulations of CPW and microstrip antenna excitation spectra}
We used CST microwave studio software to simulate the spin-wave excitation spectra of the different antenna structures \cite {Wei2008}. This commercial solver of Maxwell's equations uses a finite integration method to calculate the rf magnetic field $\mu_{0} h^{rf}$ and its in-plane ($\mu_{0} h^{rf}_{x}$, $\mu_{0} h^{rf}_{y}$) and out-of-plane ($\mu_{0} h^{rf}_{z}$) components. Since the excitation field along the CPW or antenna ($\mu_{0} h^{rf}_{x}$) is nearly uniform and $\mu_{0} h^{rf}_{z}$ is much smaller than $\mu_{0} h^{rf}_{y}$, we Fourier-transformed only the latter component. Figure \ref{Fig2} depicts several CPW and antenna configurations used in the experiments together with their simulated spin-wave excitation spectra. The large prepatterned CPW on a GaAs substrate (Fig. \ref{Fig2}(a)), which is used for VNA-FMR measurements, mainly excites spin waves with $k \approx 0$ rad/$\mu$m (Fig. \ref{Fig2}(d)). The excitation spectrum of the smaller integrated CPW with a 2-$\mu$m-wide signal line (Fig. \ref{Fig2}(b)) includes one main spin-wave mode with wave vector $k_{1}$ = 0.76 rad/$\mu$m and several high-order modes $k_{2} - k_{7}$ (Fig. \ref{Fig2}(e)). The 4-$\mu$m-wide microstrip antenna (Fig. \ref{Fig2}(c)) mainly excites spin waves with $k_1$ ranging from 0 to 1.5 rad/$\mu$m and some higher order modes at $k_2 \approx 2.0$ rad/$\mu$m and $k_3 \approx 3.8$ rad/$\mu$m (Fig. \ref{Fig2}(f)). The insets of Figs. \ref{Fig2}(d-f) show the simulated rf magnetic fields $\mu_{0} h^{rf}_{y}$ along the $y$-axis for each antenna structure.  

\section{Results}
\subsection{VNA-FMR} 
\begin{figure}[t]
    \vspace{10pt} 
    \resizebox*{0.98\columnwidth}{!}{\includegraphics{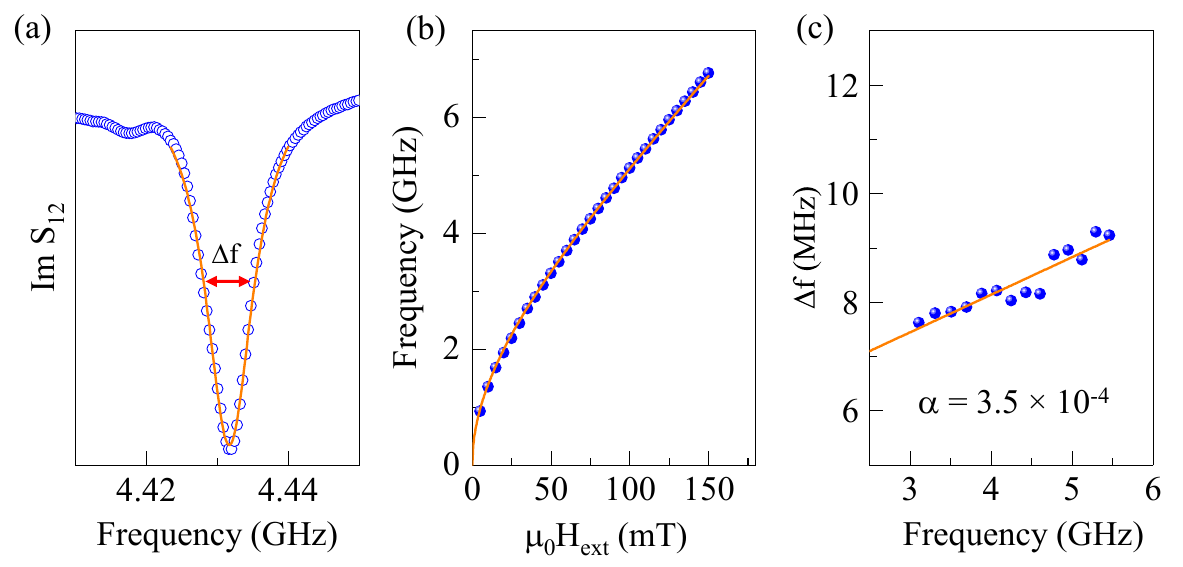}}
    \caption{(a) Imaginary part of the $S_{12}$ scattering parameter showing FMR for an in-plane external magnetic bias field of 80 mT along the CPW. The orange line is a Lorentzian function fit. (b) FMR frequency as a function of external magnetic bias field. The orange line represents a fit to the experimental data using the Kittel formula. (c) Dependence of FMR linewidth ($\Delta f$) on resonance frequency. From a linear fit to the data, we derive $\alpha$ = $3.5 \times 10^{-4}$.}
    \label{Fig3}
\end{figure}

We recorded FMR spectra for various in-plane external magnetic bias fields by measuring the $S_{12}$ scattering parameter on a 40-nm-thick YIG film. As an example, the imaginary part of $S_{12}$ recorded with a magnetic bias field $\mu_{0} H_{ext} = 80$ mT is shown in Fig. \ref{Fig3}(a). The spectrum was subtracted a reference measured at a bias field of 200 mT for enhancing signal-to-noise ratio. The resonance at $f = 4.432$ GHz is fitted by a Lorentzian function. From similar data taken at other bias fields, we extracted the field-dependence of FMR frequency and the evolution of resonance linewidth ($\Delta f$) with frequency. Figures \ref{Fig3}(b) and \ref{Fig3}(c) summarize our results. Fitting the data of Fig. \ref{Fig3}(b) to the Kittel formula $f_{res} = \frac{\gamma \mu_{0}}{2\pi} \sqrt{H_{ext}(H_{ext} + M_{eff})}$ using $\gamma /2 \pi=28$ GHz/T, we find $M_{eff}=184$ kA/m. The measured value of $M_{eff}$ is comparable to those of other PLD-grown YIG thin films \cite {Sokolov2016, Krysztofik2017}, but it is large compared to $M_{s}$ (115 kA/m). Since $M_{eff}$ = $M_{s}$ - $H_{ani}$, this means that the anisotropy field $H_{ani}=-69$ kA/m in our film. The negative anisotropy field is caused by a lattice mismatch between the YIG film and GGG substrate \cite {Sokolov2016}. Fitting the data of Fig. \ref{Fig3}(c) using $\Delta f = 2 \alpha f + \upsilon_{g} \Delta k$ gives a Gilbert damping constant $\alpha$ = $3.5 \times 10^{-4}$, which is comparable to other experiments on PLD-grown films \cite {Sun2012, Kelly2013, Howe2015}. In the fitting formula, $\upsilon_{g}$ and $\Delta k$ are the spin-wave group velocity and excitation-spectrum width, respectively \cite {Vlaminck2010}.

\begin{figure*}[t]
	\vspace{10pt} 
	\resizebox*{1.8\columnwidth}{!}{\includegraphics{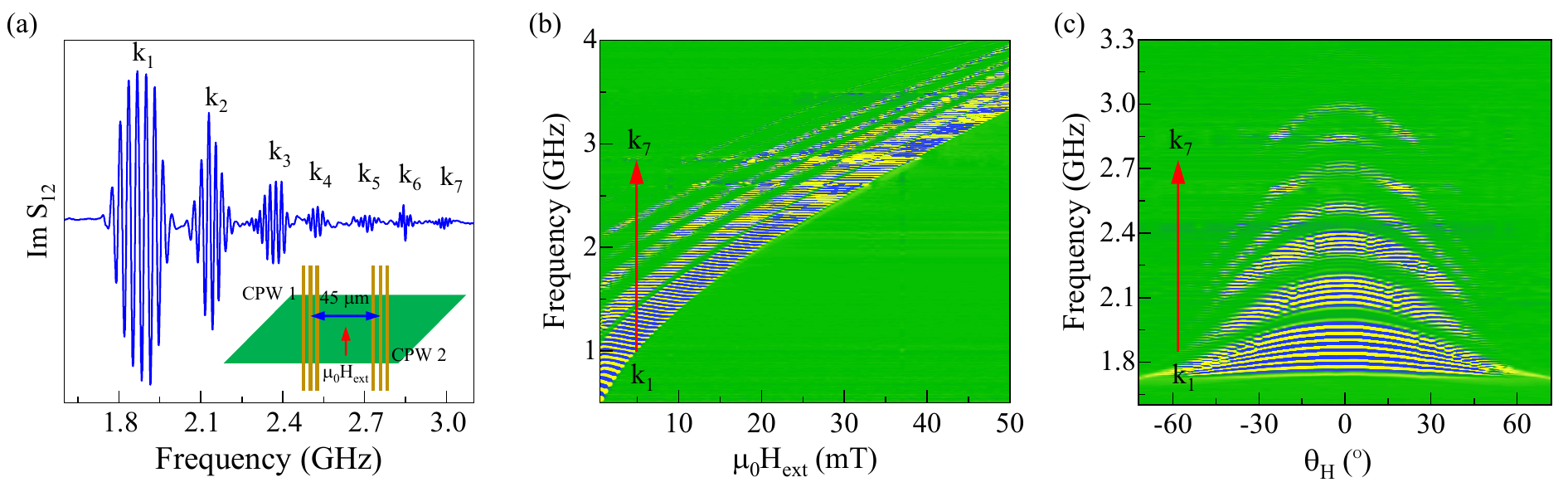}}	\caption{(a) Spin-wave transmission spectrum (imaginary part of $S_{12}$) recorded on a 40-nm-thick YIG waveguide with an external magnetic bias field $\mu_{0}H_{ext}=15.5$ mT along the CPWs. The inset shows a top-view schematic of the measurement geometry. (b) 2D map of spin-wave transmission spectra measured as a function of magnetic bias field strength. (c) Angular dependence of spin-wave transmission spectra for a constant bias field of 15.5 mT. The field angle $\theta_{H}=0^{\circ}$ corresponds to the Damon-Eshbach configuration.}
	\label{Fig4}
\end{figure*}

\subsection{Propagating spin waves}
We measured spin-wave transmission spectra on a 40-nm-thick YIG film. The measurement geometry consisted of two CPWs on top of YIG waveguides with 45$^\circ$ edges (see the inset of Fig. \ref{Fig4}(a)). The CPW parameters were identical to those in Fig. \ref{Fig2}(b) and their signal lines were separated by 45 $\mu$m. During broadband spin-wave spectroscopy, spin waves with characteristic wave vectors $k_{i}$ ($i$ = 1, 2...) were excited by passing a rf current through one of the CPWs. After propagation through the YIG film, the other CPW inductively detected the spin waves. Figure \ref{Fig4}(a) shows the imaginary part of the $S_{12}$ scattering parameter for an external magnetic bias field $\mu_{0}H_{ext}=15.5$ mT parallel to the CPWs (Damon-Eshbach configuration). The graph contains seven envelope-type peaks ($k_{1} - k_{7}$) with clear periodic oscillations. The peak intensities decrease with frequency because of reductions in the excitation efficiency and spin-wave decay length. The oscillations signify spin-wave propagation between the CPWs \cite {Vlaminck2010}. Figure \ref{Fig4}(b) shows a 2D representation of spin-wave transmission spectra recorded at different bias fields. As the field strengthens, the frequency gaps between spin-wave modes become smaller. Figure \ref{Fig4}(c) depicts the angular dependence of $S_{12}$ spectra at a constant magnetic bias field of 15.5 mT. In this measurement, the in-plane magnetic bias field was rotated from -72$^{\circ}$ to 72$^{\circ}$, where $\theta_{H} = 0^{\circ}$ corresponds to the Damon-Eshbach configuration. As the magnetization rotates towards the wave vector of propagating spin waves, the frequency and intensity of the $k_{1} - k_{7}$ modes drop. The frequency evolutions of the spin-wave modes in Figs. \ref{Fig4}(b) and \ref{Fig4}(c) are explained by a flattening of the dispersion relation with increasing magnetic bias field strength and angle. 

\begin{figure}[t]
	\vspace{10pt} 
	\resizebox*{0.9\columnwidth}{!}{\includegraphics{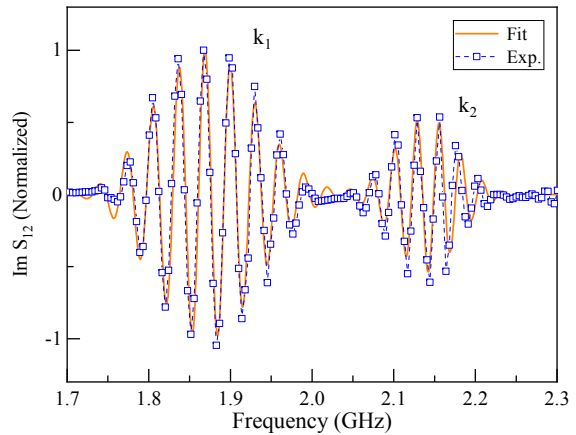}}
	\caption{A fit to the spectrum for $\mu_{0} H_{ext}=15.5$ mT and $\theta_{H}=0^{\circ}$ (blue squares) using Eq. \ref{Eq:S12-fit} (orange line).}
	\label{Fig5}
\end{figure}
 
\section{Discussion}
\subsection{Fitting of spin-wave transmission spectra}
We used Eq. \ref{Eq:PSW} to fit spin-wave transmission spectra. In this equation, $\chi (\omega, k)$ is described by a Lorentzian function, while the excitation spectrum $|\rho (k)|^{2}$ is approximated by a Gaussian function (see Fig. \ref{Fig2}(e)). For Damon-Eshbach spin waves with $kd \ll 1$, the wave vector is given by $k = \frac {2}{d} \frac{(2 \pi f)^2 - (2 \pi f_{res})^2}{(\gamma \mu_{0} M_{s})^2}$, where $d$ is the film thickness. Based on these approximations, we rewrite Eq. \ref{Eq:PSW} as: 
\begin{equation}\label{Eq:S12-fit}
\begin{aligned}
Im S_{12} \propto \frac {\Delta f}{(f-f_{res})^2+(\Delta f)^2} \times e^{-4ln2 (k - k_{0})^2/ \Delta k^2} \\
\times \sin(ks + \Phi),
\end{aligned}
\end{equation}   
where $\Delta f$ is the $S_{12}$ envelope width, $\Delta k$ is the width of the spin-wave excitation spectrum, $\Phi$ is the initial phase, and $s$ is the propagation distance. Figure \ref{Fig5} shows a fitting result for a spin-wave transmission spectrum with $\mu_{0} H_{ext}$ = 15.5 mT and $\theta_{H} = 0^{\circ}$. As input parameters, we used $f_{res}$ = 1.75 GHz, $d$ = 40 nm, $s$ = 45 $\mu$m, and $M_{eff}$ = 184 kA/m, which are either determined by geometry or extracted from measurements. $\Delta f$, $\Delta k$, $k_{0}$ are fitting parameters. For the $k_{1}$ peak, we obtained the best fit for $\Delta f$ = 0.25 GHz, $\Delta k$ = 0.6 rad/$\mu$m, and $k_{1}=0.72$ rad/$\mu$m. The $k_{2}$ peak was fitted with $k_{2}=1.87$ rad/$\mu$m. The values of $\Delta k$, $k_{1}$, and $k_{2}$ are in good agreement with the simulated excitation spectrum of the CPW (Fig. \ref{Fig2}(e)) and $\Delta f$ matches the width of the envelope peak in the experimental $S_{12}$ spectrum.      

\subsection{Spin-wave dispersion relations}
\begin{figure}[t]
	\vspace{10pt} 
	\resizebox*{0.98\columnwidth}{!}{\includegraphics{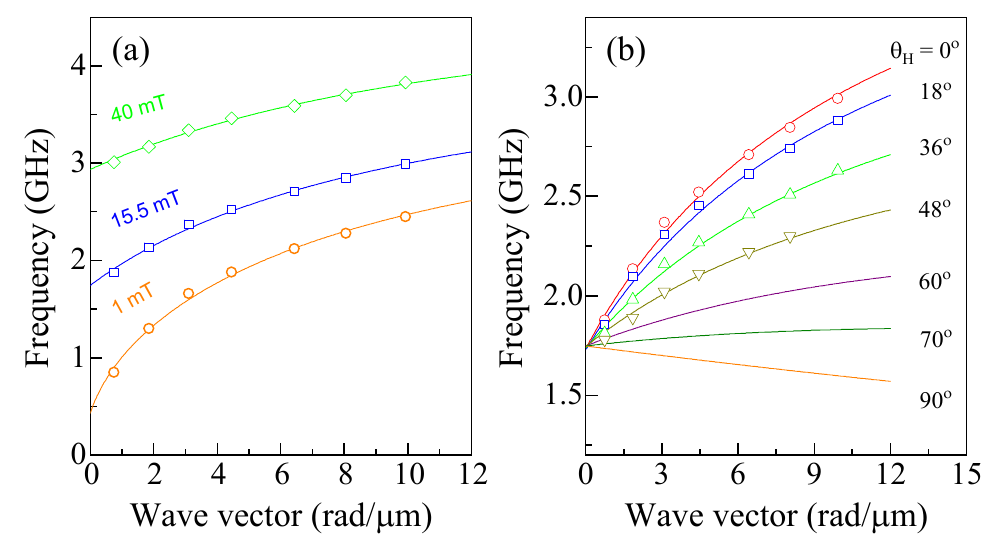}}
	\caption{Spin-wave dispersion relations for different external magnetic bias fields (a) and field angles (b). In (a) $\theta_{H} = 0^{\circ}$ and in (b) $\mu_{0} H_{ext}$ = 15.5 mT. The colored lines represent fits to the disperion relations using Eq. \ref {Eq:KS equation}.}
	\label{Fig6}
\end{figure}

We extracted spin-wave dispersion relations for different magnetic bias fields and field angles by fitting the $S_{12}$ transmission spectra shown in Figs. \ref{Fig4}(b) and \ref{Fig4}(c). The symbols in Fig. \ref{Fig6} summarize the results. We also calculated the dispersion relations using the Kalinikos and Slavin formula \cite{Kalinikos1986}:
\begin{equation}\label{Eq:KS equation}
\begin{aligned}
f = \frac{\gamma \mu_{0}} {2\pi} \bigg{[} H_{ext} \big{(}H_{ext} +  M_{eff} \big {[} 1 - F\sin^{2} \theta_{H} \\ 
+ \frac {M_{eff}} {H_{ext}} F (1 - F) \cos^{2}\theta_{H}\big{]}\big{)}\bigg{]}^{1/2},
\end{aligned}
\end{equation} 
with $F = 1 - \frac {1 - exp(-kd)}{kd}$. The calculated dispersion relations for $\gamma/2\pi$ = 28 GHz/T, $M_{eff}$ = 184 kA/m, and $d$ = 40 nm are shown as lines in Fig. \ref{Fig6}. 

\begin{figure}[t]
	\vspace{10pt} 
	\resizebox*{0.98\columnwidth}{!}{\includegraphics{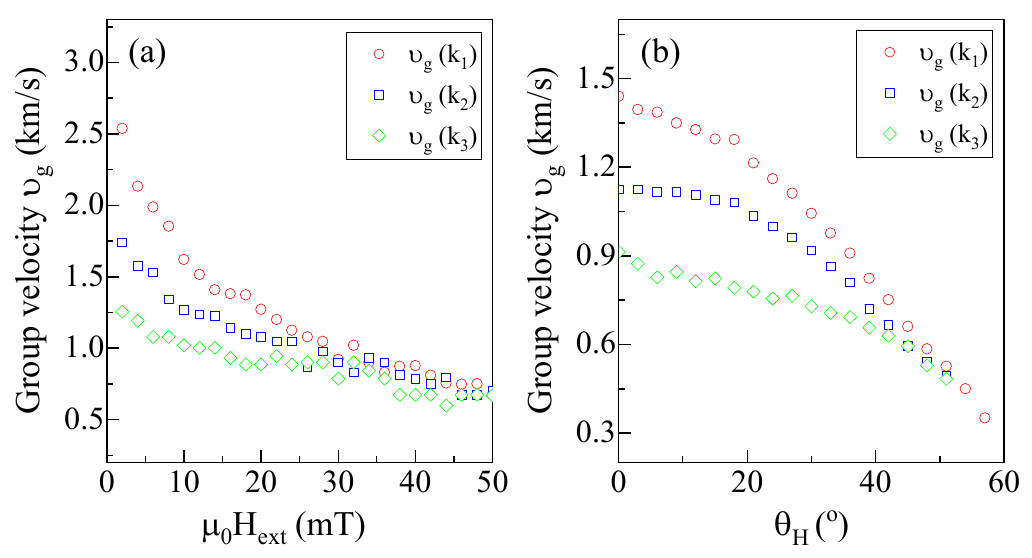}}
	\caption{Spin-wave group velocity $\upsilon_{g}$ of $k_{1} - k_{3}$ modes as a function of external magnetic bias field (a) and field angle (b). In (a) $\theta_{H} = 0^{\circ}$ and in (b) $\mu_{0} H_{ext}$ = 15.5 mT.}
	\label{Fig7}
\end{figure}

The dispersion curves flatten with increasing magnetic bias field. For instance, at $\mu_{0}H_{ext}$ = 1 mT, the frequency of propagating spin waves changes from 0.5 GHz to 2.4 GHz for wave vectors ranging from 0 to 10 rad/$\mu$m. At $\mu_{0}H_{ext}$ = 40 mT, the frequency evolution with wave vector is reduced to $3-3.7$ GHz. This magnetic-field dependence of the dispersion relation narrows the spin-wave transmission bands in Fig. \ref{Fig4}(b) at large $\mu_{0}H_{ext}$. 

The angular dependence of the spin-wave dispersion curves in Fig. \ref{Fig6}(b) is explained by in-plane magnetization rotation from $M \perp k$ ($\theta_{H} = 0^{\circ}$) towards $M \parallel k$ ($\theta_{H} = 90^{\circ}$). At $\theta_{H} = 0^{\circ}$, dispersive Damon-Eshbach spin waves with positive group velocity propagate between the CPWs. The character of excited spin waves changes gradually with increasing $\theta_{H}$ until it has fully transformed into a backward-volume magnetostatic mode at $\theta_{H} = 90^{\circ}$. This mode is only weakly dispersive and exhibits a negative group velocity.   

\subsection {Group velocity}
The phase relation between signals from the two CPWs is given by $\Phi = k \times s$ \cite{Yu2014, Vlaminck2010}. Since the phase shift between two neighboring maxima ($\delta f$) in broadband spin-wave transmission spectra corresponds to 2$\pi$, the group velocity can be written as:
\begin{equation}\label{Eq:velocity}
\upsilon_{g} = \frac {\partial \omega}{\partial k} = \frac {2\pi \delta f} {2 \pi/s} = \delta f \times s,
\end{equation} 
where $s$ = 45 $\mu$m in our experiments. Using this equation, we extracted the spin-wave group velocity for wave vectors $k_{1} - k_{3}$ from the transmission spectra shown in Figs. \ref{Fig4}(b) and \ref{Fig4}(c). Figure \ref{Fig7} summarizes the variation of $\upsilon_{g}$ with external magnetic bias field and field angle. For weak bias fields ($\mu_{0}H_{ext}<30$ mT), the group velocity decreases swiftly, especially if $k$ is small. For instance, $\upsilon_{g}(k_{1})$ reduces from 3.0 km/s to 1.0 km/s in the $0-30$ mT field range, while $\upsilon_{g}(k_{3})$ only changes from 1.2 km/s to 0.8 km/s. At larger external magnetic bias fields, $\upsilon_{g}$ decreases more slowly for all wave vectors. Figure \ref{Fig7}(b) shows how $\upsilon_{g}$ varies as a function of field angle at $\mu_{0}H_{ext}$ = 15.5 mT. For all wave vectors, the group velocity is largest in the Damon-Eshbach configuration ($\theta_{H}=0^\circ$). At larger field angles, $\upsilon_{g}$ decreases and its dependence on wave vector diminishes. Variations of the spin-wave group velocity with wave vector and magnetic-field strength or angle are explained by a flattening of the dispersion relations, as illustrated by the data in Fig. \ref{Fig6}.

\begin{figure}[t]
	\vspace{10pt} 
	\resizebox*{0.98\columnwidth}{!}{\includegraphics{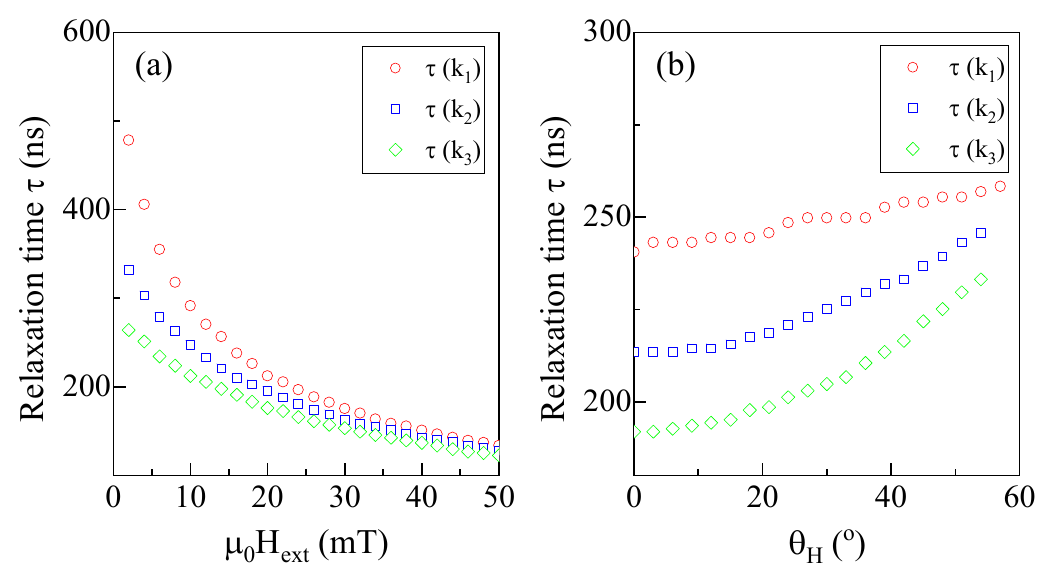}}
	\caption{Spin-wave relaxation time $\tau$ of $k_{1} - k_{3}$ modes as a function of external magnetic bias field (a) and field angle (b). In (a) $\theta_{H} = 0^{\circ}$ and in (b) $\mu_{0} H_{ext}$ = 15.5 mT.}
	\label{Fig8}
\end{figure}

\begin{figure}[t]
	\vspace{10pt} 
	\resizebox*{0.98\columnwidth}{!}{\includegraphics{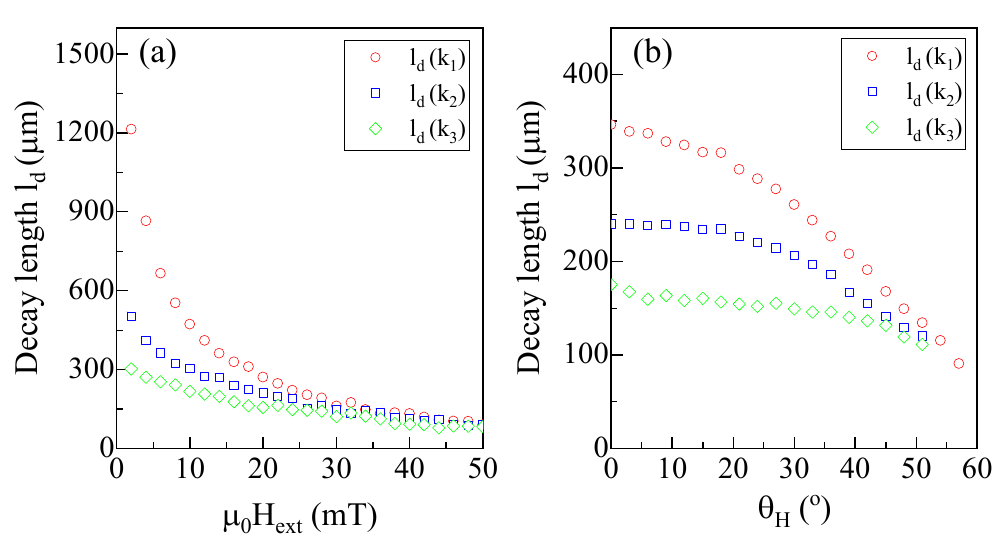}}
	\caption{Spin-wave decay length $l_{d}$ of $k_{1} - k_{3}$ modes as a function of external magnetic bias field (a) and field angle (b). In (a) $\theta_{H} = 0^{\circ}$ and in (b) $\mu_{0} H_{ext}$ = 15.5 mT.}
	\label{Fig9}
\end{figure} 

\begin{figure*}[t]
	\vspace{10pt} 
	\resizebox*{1.8\columnwidth}{!}{\includegraphics{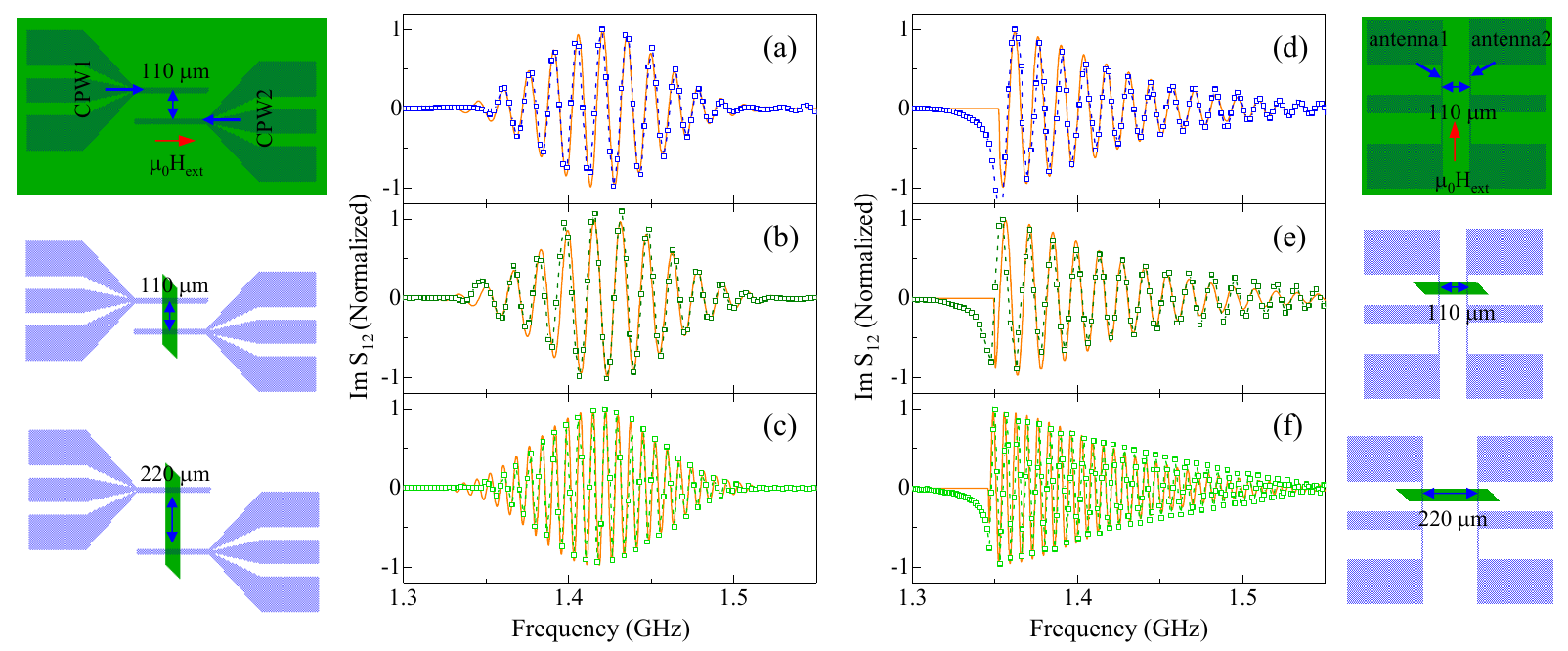}}
	\caption{(a)-(c) Spin-wave transmission spectra measured using CPWs on a continuous YIG film (a) and 50-$\mu$m-wide YIG waveguides ((b) and (c)). The YIG film and waveguides are 35 nm thick and the CPWs are separated by 110 $\mu$m ((a) and (b)) and 220 $\mu$m (c). (d)-(f) Spin-wave transmission spectra measured using microstrip antennas on the same YIG film and waveguides. The signal lines of the CPWs and microstrip antennas are 4 $\mu$m wide. The orange lines represent fits to the experimental data using Eq. \ref{Eq:S12-fit}. The measurement geometry for each spectrum is illustrated next to the graphs. In the schematics, the green areas depict a continuous YIG film or waveguide.}
	\label{Fig10}
\end{figure*}

\subsection {Spin-wave relaxation time and decay length}
We now discuss the relaxation time ($\tau$) and decay length ($l_{d}$) of spin waves in our YIG films. Following Ref. \citenum{Madami2011}, the relaxation time is estimated by $\tau = 1/ 2\pi \alpha f$. Using $\alpha=3.5\times10^{-4}$ and spin-wave transmission data from Fig. \ref{Fig4}, we determined $\tau$ for wave vectors $k_{1} - k_{3}$. The dependence of $\tau$ on external magnetic bias field and field angle is shown in Fig. \ref{Fig8}. The maximum spin-wave relaxation time in our 40-nm-thick YIG films is approximately 500 ns. Resembling the spin-wave group velocity, $\tau$ is largest for small wave vectors and it decreases with increasing bias field (Fig. \ref{Fig8}(a)). In contrast to $\upsilon_{g}$, the spin-wave relaxation time is smallest in the Damon-Eshbach configuration ($\theta_{H}=0^\circ$) and it evolves more strongly with increasing $\theta_{H}$ if $k$ is large (Fig. \ref{Fig8}(b)). This result is explained by $\tau\propto1/f$ and a lowering of the spin-wave frequency if the in-plane bias field rotates the magnetization towards $k$ (see Fig. \ref{Fig4}(c)). 

The spin-wave decay length is calculated using $l_{d} = \upsilon_{g} \times \tau$ and data from Figs. \ref{Fig7} and \ref{Fig8}. Figure \ref{Fig9}(a) shows the dependence of $l_{d}$ on $\mu_{0} H_{ext}$ for wave vectors $k_{1} - k_{3}$. The largest spin-wave decay length in our 40-nm-thick YIG films is 1.2 mm, which we measured for $k_{1}=0.72$ rad/$\mu$m and $\mu_{0} H_{ext}=2$ mT. The decay length decreases with magnetic bias field to about 100 $\mu$m at $\mu_{0} H_{ext}=50$ mT. Figure \ref{Fig9}(b) depicts the dependence of $l_{d}$ on the direction of a 15.5 mT bias field. The spin-wave decay length decreases substantially with $\theta_{H}$ for small $k$, but its angular dependence weakens for larger wave vectors. 

The decay of propagating spin waves between the exciting and detecting CPW in the broadband spectroscopy measurement geometry is given by $\exp(-s/l_{d})$ \cite {Madami2011}. Based on the results of Fig. \ref{Fig9}, one would thus expect the intensity of spin waves to drop with increasing wave vector and in-plane bias field strength or angle. The spin-wave transmission spectra of Fig. \ref {Fig4} confirm this behavior. 

\subsection{CPWs versus microstrip antennas}
Finally, we compare broadband spin-wave spectroscopy measurements on YIG thin films using CPWs and microstrip antennas. In these experiments, the CPWs and antenna structures have 4-$\mu$m-wide signal lines and they were patterned onto the same 35-nm-thick YIG film. For comparison, we also recorded transmission spectra on 50-$\mu$m wide YIG waveguides. The separation distance ($s$) between the CPWs or microstrip antennas was set to 110 $\mu$m or 220 $\mu$m. Schematics of the different measurement geometries are depicted on the sides of Fig. \ref{Fig10}. Transmission spectra that were obtained for Damon-Eshbach spin waves in each configuration are also shown. In all measurements, we used an in-plane external magnetic bias field of 10 mT. The plots focus on phase oscillations in the first-order excitation at $k_{1}$ (higher-order excitations were measured also, but are not shown). The differently shaped outline of the $S_{12}$ peak for two CPWs (left) or two microstrip antennas (right) mimics the profile of their excitation spectra (Fig. \ref{Fig2}). As expected from $\delta f=\upsilon_{g}/s$, the period of frequency oscillations ($\delta f$) becomes smaller if the separation between antennas ($s$) is enhanced (Figs. \ref{Fig10}(c) and \ref{Fig10}(f)).       

We fitted the spin-wave transmission spectra obtained with CPWs (Figs. \ref{Fig10}(a)-(c)) using the same procedure as described earlier. Good agreements between experimental data (blue squares) and calculations (orange lines) were obtained by inserting $M_{eff}$ = 190 $\pm$ 4 kA/m, $\Delta f$ = 0.18 GHz, $k$ = 0.34 rad/$\mu$m, and $\Delta k$ = 0.33 rad/$\mu$m into Eq. \ref{Eq:S12-fit}. To fit $S_{12}$ spectra measured by microstrip antennas, we approximated the wave vector of the excitation as $k = \frac {2}{d} \frac{(2\pi f)^2-(2 \pi f_{res})^2}{(\gamma \mu_{0} M_{eff})^2} H(f - f_{res})$, where $H$ is a Heaviside step function \cite{Ciubotaru2016}. The best results were achieved for $M_{eff}$ = 178 $\pm$ 2 kA/m, $\Delta f$ = 0.25 GHz, $k$ = 0 rad/$\mu$m and $\Delta k$ = 0.65 rad/$\mu$m. From the data comparison in Fig. \ref{Fig10}, we conclude that broadband spin-wave spectroscopy measurements with CPWs and microstrip antennas yield similar results for $M_{eff}$. We also note that the $S_{12}$ peak width ($\Delta f$) obtained from measurements on continuous YIG films and YIG waveguides are nearly identical ($\Delta f = 0.18$ GHz for CPWs, $\Delta f = 0.22$ GHz for antennas). Thus, patterning of the YIG film into waveguides does not deteriorate the Gilbert damping constant.   

From the oscillation periods ($\delta f$) in the transmission spectra of Fig. \ref{Fig10}, we extracted the properties of propagating spin waves. By averaging $\delta f$ over the same frequency range in spectra measured by CPWs and microstrip antennas, we obtained $\upsilon_{g}=1.67$ km/s and $\upsilon_{g} = 1.53$ km/s, respectively. The spin-wave relaxation time was determined as $\tau = 225$ ns (CPW) and $\tau=237$ ns (antenna) and the decay length was extracted as $l_{d} = 375$ $\mu$m (CPW) and $l_{d} = 363$ $\mu$m (antenna). These results clearly demonstrate that broadband spin-wave spectroscopy measurements on YIG thin films using CPWs or microstrip antennas provide comparable results.

\section{Summary} 
In conclusion, we prepared $35-40$ nm thick epitaxial YIG films with a Gilbert damping constant $\alpha$ = $3.5 \times 10^{-4}$ on GGG(111) substrates using PLD. The dependence of spin-wave transmission on the strength and angle of an in-plane magnetic bias field was systematically gauged. We used the measurements to demonstrate strong tuning of the spin-wave group velocity ($\upsilon_{g}$), relaxation time ($\tau$), and decay length ($l_{d}$) up to a field strength of about 30 mT and above a field angle of 20$^{\circ}$. Maximum values of $\upsilon_{g} = 3.0$ km/s, $\tau = 500$ ns, and $l_{d} = 1.2$ mm were extracted for Damon-Eshbach spin waves with $k_{1}$ = 0.72 rad/$\mu$m. Moreover, we demonstrated that broadband spin-wave spectroscopy performed with integrated CPWs and microstrip antennas yield similar results.

\section{Acknowledgements} 
This work was supported by the European Research Council (Grant Nos. ERC-2012-StG 307502-E-CONTROL and ERC-PoC-2018 812841-POWERSPIN). S.J.H. acknowledges financial support from the V\"ais\"al\"a Foundation. Lithography was performed at the Micronova Nanofabrication Centre, supported by Aalto University. We also acknowledge the computational resources
provided by the Aalto Science-IT project.

\end{document}